%% file: Marigo.tex
\documentclass[universe,article,accept,oneauthor,pdftex]{Definitions/mdpi} 
\newcommand{\Msun}{\mbox{$\mathrm{M}_{\odot}$}}
\newcommand{\Lsun}{\mbox{$\mathrm{L}_{\odot}$}}
\newcommand{\Mi}{\mbox{$M_{\rm i}$}}
\newcommand{\Menv}{\mbox{$M_{\rm env}$}}
\newcommand{\Mf}{\mbox{$M_{\rm f}$}}
\newcommand{\Mc}{\mbox{$M_{\rm c}$}}
\newcommand{\Mcmin}{\mbox{$M_{\rm c}^{\rm min}$}}
\newcommand{\Mdup}{\mbox{$\Delta M_{\rm 3DU}$}}

\newcommand{\co}{\mbox{${\rm C/O}$}}
\newcommand{\cminuso}{\mbox{$\mathrm{C}-\mathrm{O}$}}


\firstpage{1} 
\pubvolume{1}
\issuenum{1}
\articlenumber{0}
\pubyear{2022}
\copyrightyear{2022}
\externaleditor{Academic Editor: Sara Palmerini} 
\datereceived{} 
\dateaccepted{} 
\datepublished{} 
\hreflink{https://doi.org/} 

\usepackage{aas_macros}

\Title{The Initial-Final Mass Relation of White Dwarfs: A~Tool to Calibrate the Third Dredge-Up}

\TitleCitation{The Initial-Final Mass Relation of White Dwarfs: A Tool to Calibrate the Third Dredge-Up}

\Author{Paola Marigo\orcidA{0000-0002-9137-0773}}

\AuthorNames{Paola Marigo}

\AuthorCitation{Marigo, P.}

\address[1]{%
Department of Physics and Astronomy, University of Padova, 35122-Padova, Italy; paola.marigo@unipd.it}

\abstract{The initial mass-final mass relationship (IFMR) of white dwarfs (WD) represents a crucial benchmark for stellar evolution models, especially for the efficiency of mixing episodes and mass loss during the asymptotic giant branch (AGB) phase.
In this study, we argue that this relation offers the opportunity to constrain  the third dredge-up (3DU), with important consequences for chemical yields. The results are discussed in light of recent studies that have identified a kink in the IFMR for initial masses close to $2\,\Msun$. Adopting a physically-sound approach in which the efficiency $\lambda$ of the 3DU varies as a function of core and envelope masses, we calibrate $\lambda$ in solar-metallicity TP-AGB models in order to reproduce the final masses of their WD progeny, over the range of initial masses $0.9 \le \Mi/\Msun \le 6$. In particular, we find that in low-mass stars with $1.4 \lesssim \Mi/\Msun \lesssim 2.0$ the efficiency is small, $\lambda \le 0.3$, it steeply rises to about $\lambda \simeq 0.65$ in intermediate-mass stars with $2.0 \le \Mi/\Msun \le 4.0$, and then it drops in massive TP-AGB stars with $4.0 \lesssim \Mi/\Msun \lesssim 6.0$. Our~study also suggests that a second kink may show up in the IFMR at the transition between the most massive carbon stars and those that are dominated by hot-bottom burning.}


\begin{document}
\section{Introduction}
The initial-final mass relation (IFMR) of white dwarfs plays a key role in several astrophysical applications \citep{Marigo_13, Salaris_etal_09}.
A recent  analysis \citep{Marigo_etal_20} of a few carbon-oxygen white dwarfs in old open clusters of the Milky Way (MW) identified a kink in the initial-final mass relation (IFMR), located over a range of initial masses, $1.65 \lesssim M_{\rm i}/M_{\odot} \lesssim 2.10$, which unexpectedly interrupts the commonly assumed monotonic trend. The proposed interpretation links this observational fact to the formation of carbon stars and the modest outflows (with mass loss rate $< 10^{-7}\, M_{\odot}/\mathrm{yr}$) that are expected as long as the carbon excess, \cminuso, remains too low to produce carbonaceous dust grains in a sufficient amount.  Under these conditions the mass of the carbon-oxygen core  can grow more than is generally predicted by stellar models. 
A new systematic follow-up investigation \citep[][to which we refer for all details]{Marigo_etal_22}, based on GAIA DR2 and EDR3, of TP-AGB stars belonging to open clusters of known age support the findings of \citep{Marigo_etal_20}.

The key point in explaining the kink is the interplay between the third dredge-up (3DU), which determines the surface enrichment in carbon, and the dependence of stellar winds for C stars on the excess of carbon with respect to oxygen, \cminuso\ \citep{Mattsson_etal_10, Eriksson_etal_14, Bladh_etal_19_C}.
We leverage this mutual interconnection to constrain the efficiency of 3DU, $\lambda$,  in solar metallicity TP-AGB stars, using semi-empirical IFMR as the primary calibrator.
A similar approach was already  introduced by \citet{Kalirai_etal_14} and \citet{Marigo_etal_20}. In this work we improve the methodology, as we take into account the dependence of $\lambda$ on main stellar parameters (core mass and envelope mass), on the basis of the indications of complete TP-AGB models~\citep{Straniero_etal_03, Cristallo_etal_15}.

The paper is structured as follows.
Section~\ref{sec_methods} details the novel approach to calibrating $\lambda$ and recall the basic ingredients of our \texttt{COLIBRI} code for the TP-AGB phase.
In \mbox{Section~\ref{sec_res}} we analyze and discuss the results, with a particular focus on the impact of carbon enrichment on the mass-loss rate (Section~\ref{ssec_models}). We present our derived $\lambda$-law at solar metallicity in Section~\ref{ssec_lambda}, and compare it with the predictions of full TP-AGB models in the literature.
\mbox{Section~\ref{ssec_ifmr}} analyses the semi-empirical IFMR, compares the data with models, and points to a possible secondary kink drawn by the WDs produced by high-mass AGB stars (\mbox{Section~\ref{ssec_2kink}}).
In Section~\ref{sec_tau} we discuss  other relevant quantities that derive from the $\lambda$-calibration, namely the TP-AGB lifetimes and the carbon ejecta. We compare our findings with other published models.
Finally, Section~\ref{sec_end} closes the paper.
\section{Materials and Methods}
\label{sec_methods}
The 3DU influences the IFMR in at least two ways: 1. The mass of the core, \Mc, is instantaneously reduced at each mixing episode; 2. It affects the mass-loss rate.
The higher the efficiency $\lambda$\footnote{The efficiency of a 3DU event is commonly described by the dimensionless parameter $\lambda =\Delta M_{\rm 3DU}/\Delta M_{\rm c}$, defined as the amount of dredged-up material, $\Delta M_{\rm 3DU}$, relative to the growth of the core mass, $\Delta M_{\rm c}$, during the previous inter-pulse period.} of the dredge-up episode, the greater the decrease in \Mc, equal to $\lambda \times \Delta \Mc$.
The effect on mass loss is a consequence of the variation of the surface chemical composition which induces notable changes in atmospheric opacity \citep{MarigoAringer_09, Marigo_02} and hence in the effective temperature \citep{Marigo_etal13}, a key parameter for the pulsation periods \citep{Trabucchi_etal_19}, and the winds of AGB stars \citep{HoefnerOlofsson_18, Bladh_etal_19_M}.
Furthermore, the increase of C and O in the atmospheres of carbon stars controls the excess of carbon relative to oxygen, \cminuso, which in turn sets the budget of C atoms available for the formation of carbonaceous dust
\citep{Bladh_etal_19_C, Mattsson_etal_10, FerrarottiGail_06}.

In \citet{Marigo_etal_20}, we calibrated the average efficiency of the 3DU as a function of the initial mass to reproduce the IFMR 
and its kink at $\Mi\simeq 2\,\Msun$, under the simple hypothesis that $\lambda$ remains constant during the TP-AGB phase, at given \Mi. In this study we aim to improve the description of the 3DU, by relaxing the assumption of constant $\lambda$. In principle, the depth of the 3DU depends primarily on the mass of the core, \Mc, the mass of the envelope, \Menv, and the chemical composition (e.g., \citep{Herwig_00, Straniero_etal_03}).  These parameters, in turn, affect other key quantities for the 3DU, such as, for instance, the flash-luminosity peak $L_{\rm P}$.
At this stage the envelope convection zone reaches its maximum inward
extension in mass fraction and temperature while, at the same time, the nuclear-processed material
is pushed out to its lowest temperature during the thermal-pulse cycle \citep{Wood_81}.
In principle, the larger $L_{\rm P}$, the deeper the inward envelope extension.

An important step forward to include these dependencies was already addressed by~\citep{Pastorelli_etal_19,Pastorelli_etal_20}, in their population synthesis simulations aimed at reproducing the luminosity functions of AGB stars in the Magellanic Clouds.
In this paper, we move to solar metallicities, and adopt the IFMR relation as the main calibrator. To take a more physical approach we rely on a study based on complete TP-AGB models of the FRUITY database\footnote{Full-network Repository of Updated Isotopic Tables and Yields: \url{http://fruity.oa-teramo.inaf.it/} 
}  (\citep{Straniero_etal_03}, see their Equation (3) and Figure 1).
There, the authors proposed a fitting formula that accurately reproduces their predictions for  the mass of dredged-up material, \Mdup,  at each pulse, as a function of the minimum mass of the core, \Mcmin, for the onset 3DU, the current  envelope and core masses, \Menv\ and \Mc, and the metallicity, $Z$.

We adopt the same functional form introduced by \cite{Straniero_etal_03}, with some modifications, and fit  \Mdup\ of the entire set of \texttt{FRUITY} TP-AGB models with metallicity $Z=0.014$. 
Using the original fitting relation in combination with our mass-loss prescriptions in  \texttt{COLIBRI}, we fail to reproduce the IFMR. Therefore, we opt to maintain the dependence on \Mcmin, \Mc, and \Menv, and introduce a free parameter that allows us to vary the relation to best reproduce the observational IFMR data.

In practice, we apply a corrective factor, $f_{\rm cor}$, to the \texttt{FRUITY} \Mdup\ relation, and make it vary iteratively until we find a good match between models and observations.
This procedure is carried out for every \texttt{COLIBRI} stellar model of given initial mass \Mi, calculating, for each iteration with a test $f_{\rm cor}$, the entire TP-AGB phase up to the complete ejection of the envelope. When the final \Mc\ of the model  and the measured WD  mass coincide (within a tolerance of about 0.01--0.05 $\Msun$),
the loop closes and the appropriate value of $f_{\rm cor}$ is found.
The results of the calibration are discussed in Section~\ref{sec_res}.

\subsection*{TP-AGB Models}
\label{ssec_models}
At this point it is worth mentioning here the main input prescriptions of the \texttt{COLIBRI} code, focusing on those that are particularly relevant for the present study.
For more details we refer to \cite{Marigo_etal_20,Pastorelli_etal_20,Pastorelli_etal_19,Marigo_etal13}.
\begin{itemize}  
\item \textit{Equation of state and opacity.} \texttt{COLIBRI} contains the \texttt{\AE SOPUS}\footnote{\url{http://stev.oapd.inaf.it/cgi-bin/aesopus}}
 and \texttt{Opacity Project}\footnote{\url{http://cdsweb.u-strasbg.fr/topbase/TheOP.html}} software packages \citep{MarigoAringer_09, Seaton05} as internal routines, to compute on-the-fly both the equation of state  (for $\gtrsim$  800 atomic and molecular species) and Rosseland mean opacities at temperatures $10^3 \le T/\mathrm{K} \le 10^8$. 
This allows us to consistently describe the impact on the stellar structure of variations in abundances due to nucleosynthesis and mixing.
\item \textit{Nucleosynthesis.}
Hot-bottom burning is followed with a complete nuclear network coupled to a diffusive description of convection. Nuclear reactions of the p-p chains, CNO cycles, Ne-Na and Mg-Al chains are included.
\item \textit{Mass loss.} Before the onset of dust-driven mass loss  we assume that stellar winds are sustained by low-amplitude pulsation \cite{Winters_etal_00, Bowen_88, Bedijn_88}. The resulting rates are typically low, from $\approx$$10^{-9} $--$10^{-7}$  $\Msun$ yr$^{-1}$.
In the dust-driven regime the mass-loss rate is computed with a different formalism depending on the photospheric \co\, ratio, namely: a widely-used relation (\cite{Bloecker_95}, with an efficiency parameter $\eta_{\rm B}=0.01$)
for the O-rich stages when $\co < 1$, and a routine based on state-of-the-art pulsating atmosphere models for carbon stars, when $\co >1$ \citep{Mattsson_etal_10, Eriksson_etal_14}. These models, in particular, predict that radiation-driven winds are effectively triggered by carbonaceous dust only when (i)~the amount of free carbon, \cminuso, overcomes a threshold (\cminuso)$_{\rm min}$, and (ii) suitable conditions for dust formation (in terms of stellar luminosity, effective temperature and mass) are met in the extended atmospheres.
\item \textit{Intershell composition.} A nuclear network with the main $\alpha$-capture reactions is included. The typical abundance pattern is  characterized by  $\mathrm{He/C/O =}$ 0.70--0.78/0.20--0.25/ 0.005--0.02 (abundances in mass fraction). No overshoot is applied to the bottom of the pulse-driven convection zone.
\end{itemize}
\section{Results}
\label{sec_res}
\subsection{The Impact of Carbon Enrichment on the Mass-Loss Rate}
Two key aspects characterize our \texttt{COLIBRI} models, namely:
the computation on-the-fly of gas opacities correctly coupled to the changes in the envelope chemical composition, and
the adoption of a mass-loss prescription for the carbon-rich phase that is explicitly dependent on carbon excess, \cminuso\ (see Section~\ref{ssec_models}). 
This allows us to explore the feedback due to carbon enrichment both on the stellar structure and on mass loss, and their mutual interaction 
as well.

\begin{figure}[H]
\centering
\includegraphics[width=6.5cm]{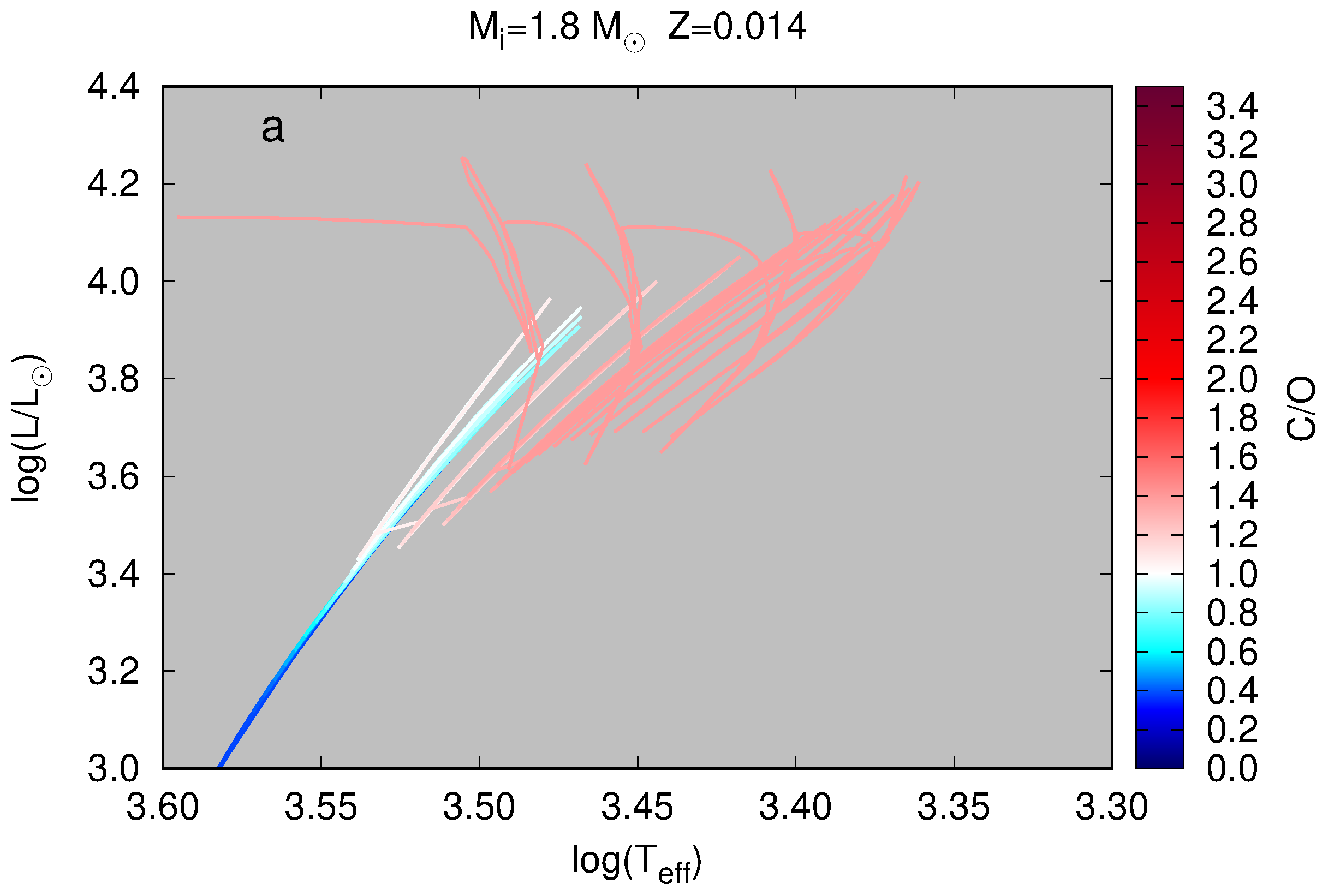}
\includegraphics[width=6.5cm]{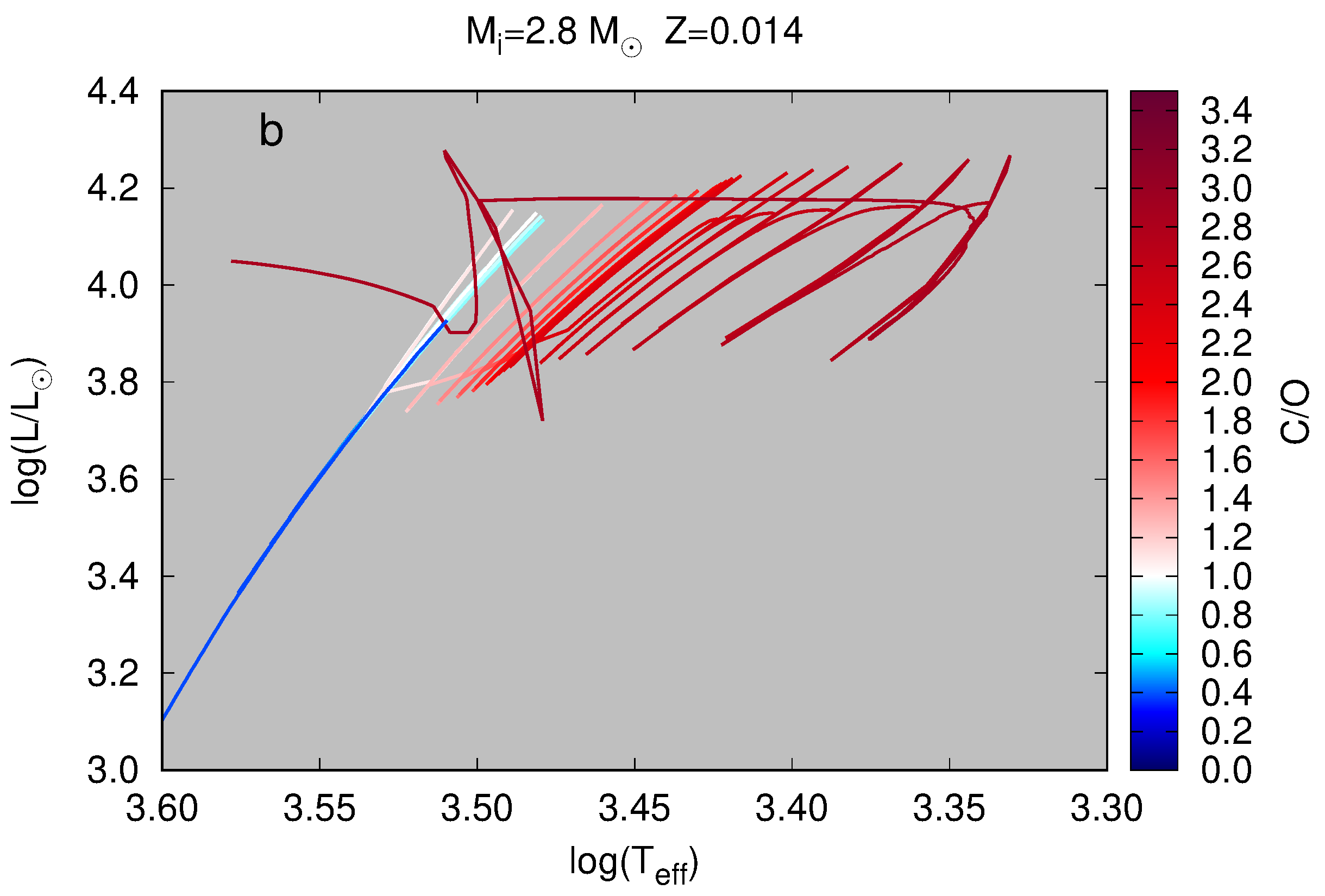}
\includegraphics[width=6.5cm]{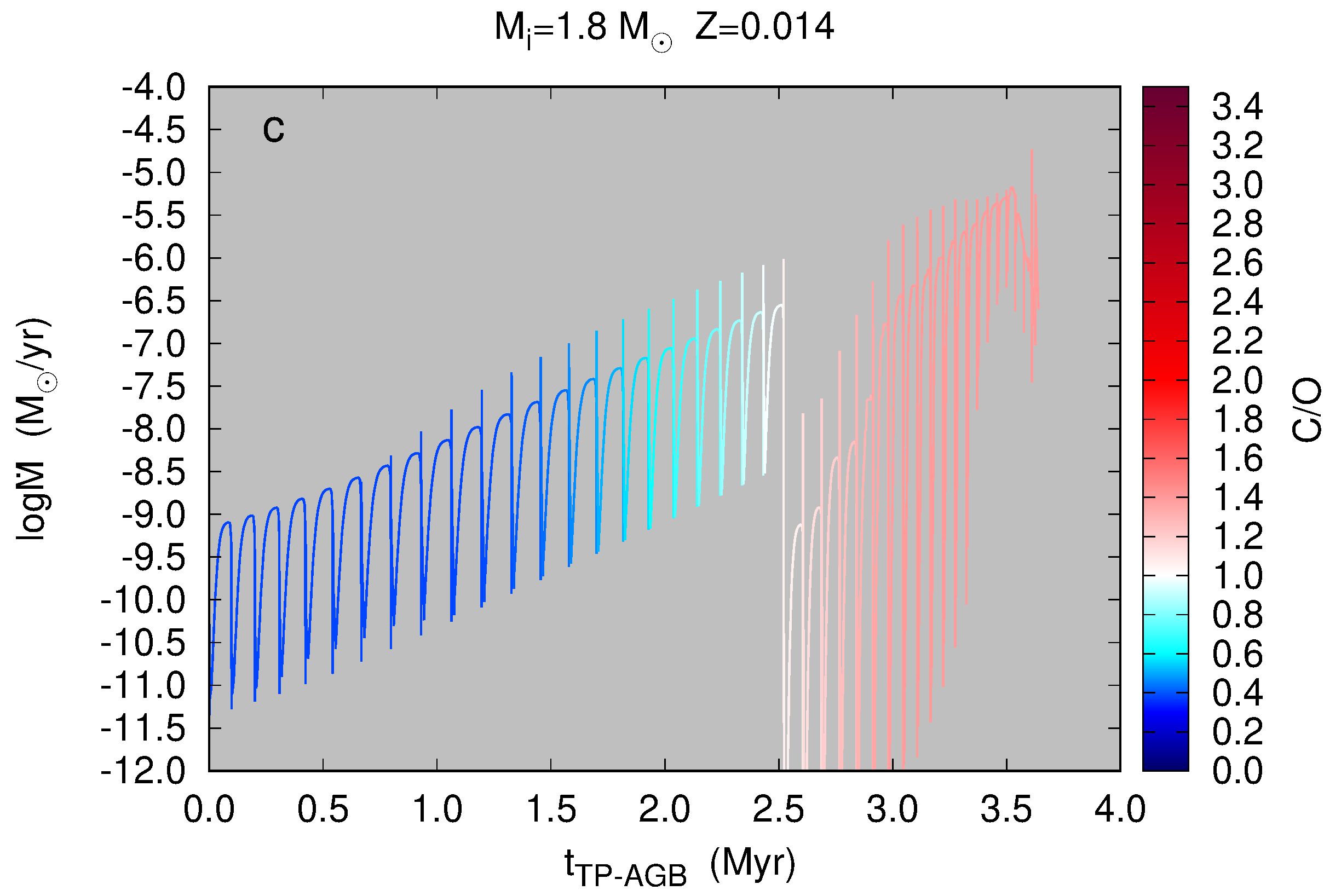}
\includegraphics[width=6.5cm]{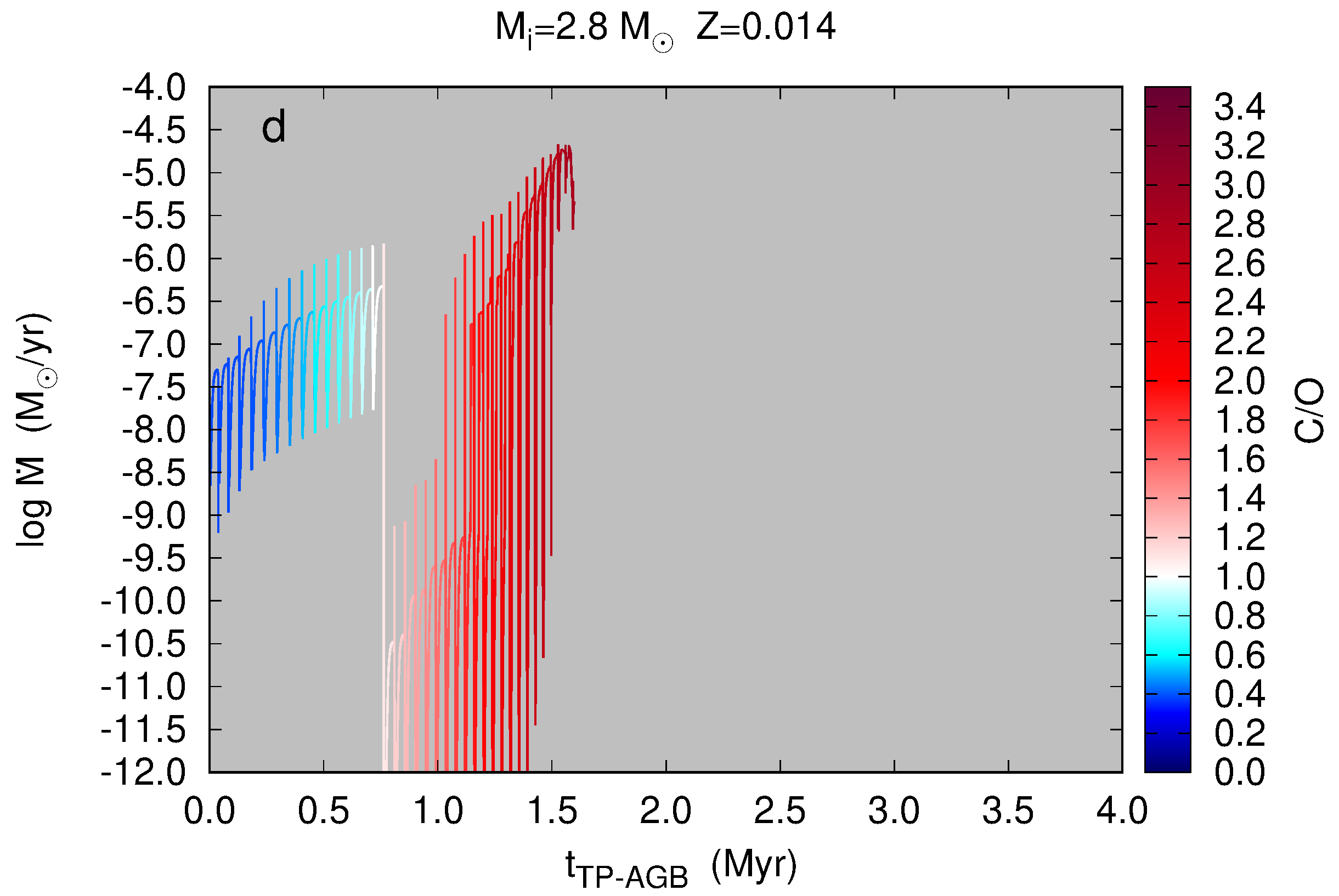}
\caption{Impact of carbon enrichment on the evolution of TP-AGB stars. Two models are shown as an example, namely $\Mi=1.8\, \Msun$ (left panels) and $\Mi=2.8\,\Msun$ (right panels), which reach approximately the same final mass, $\Mf\simeq 0.70--0.71\,\Msun$ at the end of the TP-AGB phase.
(\textbf{a},\textbf{b})~Evolution on the H-R diagram during the whole TP-AGB evolution, color-coded according to the current photospheric \co\ ratio.
(\textbf{c},\textbf{d}) Evolution of the mass-loss rate.  Time is set to zero at the first TP. The tracks are color-coded according to the current photospheric \co\ ratio. 
The drop in mass loss during the carbon-star phase occurs when the carbon excess, \cminuso, is below the current threshold to trigger a dust-driven wind. See the text for more details.
\label{fig1}}
\end{figure}

Figure~\ref{fig1} shows the evolution of luminosity, effective temperature and mass-loss rate, in two models that  become both carbon stars and end their TP-AGB phase with a bare core of about $\Mf \simeq 0.7\,\Msun$.
{First, we note that when \co\ increases due to the 3DU and enters the narrow range $0.98\lesssim \co \lesssim 1.05$ (shades from light pink to white), which characterizes the stars belonging to the advanced S and SC spectral classes, the tracks in the HR diagram tend to heat up slightly (due to a minimum in opacity, since most of C and O atoms are locked in the CO molecule; see \citet{MarigoAringer_09}). Then, when \co\ exceeds $\simeq 1.05$ and attains larger values as a consequence of repeated 3DU episodes, the atmospheric opacity increases making the effective temperature decrease sizeably, a well-known property that distinguishes the class of carbon stars \citep{Marigo_02}. We see that the model with $M_{\rm i}= 2.8\, M_{\odot}$ achieves a higher surface enrichment in carbon, and therefore the track cools more than the $M_{\rm i}= 1.8\, M_{\odot}$ model. In turn, the increase in C/O and the consequent decrease in $T_{\rm eff}$ favor the increase in mass loss (bottom panels).
In both tracks, the minimum in $T_{\rm eff}$ is reached when a large part of the envelope has been ejected, i.e., as soon as the envelope mass falls below 35--40\% of the total mass ($M_{\rm env}/M <$ 0.35--0.40). From that stage on, the tracks reverse their evolutionary direction, and begin to heat up as the last thermal pulses occur. In the present models the evolutionary calculations stop when $M_{\rm env} < 0.05\, M_{\odot}$.}

Let us now move to discuss the evolution of the mass-loss rate (bottom panels of Figure~\ref{fig1}). During the O-rich phases it gradually increases, as expected from the \cite{Bloecker_95} formula.
Then, as soon as the model passes through 
 the transition from $\co < 1$ to $\co >1$, the mass-loss rate drops and remains moderate ($\dot M < {5} \times 10^{-7}\, \Msun \mathrm{yr^{-1}}$)
 until the carbon enrichment, resulting from the 3DU, reaches and exceeds the limit necessary to activate a dust-driven wind. 

This threshold is not constant, but depends on  stellar parameters: in general it increases with the current mass of the star and its effective temperature. In other words, more massive TP-AGB stars must enrich more in carbon to form the minimum amount of carbonaceous dust necessary to trigger the radiative wind.
This is evident from the comparison of panels \textbf{c} and \textbf{d}: the model with $\Mi = 2.8\, \Msun$ must attain 
$\co \gtrsim 1.8$ to power a dust-driven wind, while in the model with $\Mi = 1.8\, \Msun$ this happens when $\co \gtrsim 1.3$. {The fact that the model with $\Mi = 2.8\, \Msun$ must exceed a higher \cminuso\ threshold explains why the dust-driven wind regime (with $\dot M \gtrsim 10^{-6}\, \Msun\,{\rm yr^{-1}}$) is activated about 13 TPs after the transition 
to the C-star regime, while in the model with $\Mi = 1.8\, \Msun$ the same mass-loss rate is reached after 9 TPs. In fact, although the efficiency of 3DU for $\Mi = 2.8\, \Msun$ is greater, the carbon surface enrichment is slower from pulse to pulse, as the dredged-up material is diluted in a more massive envelope. Overall, the model with $\Mi = 2.8\, \Msun$ achieves a higher carbon enrichment and ends the TP-AGB phase with $\co=2.84$, while the model with $\Mi = 1.8\, \Msun$ terminates its life with  $\co=1.40$.}

In both cases the main consequences are: 1. A transient period of modest mass loss during the C-rich phase in which the star evolves along its Hayashi line at approximately constant mass and without significant dust production; 2. A prolongation of the TP-AGB lifetime compared to the predictions obtained with mass loss formulas that do not depend on the surface carbon excess.
In particular, for the model with $\Mi = 1.8\,\Msun$, this circumstance  allows the core to grow more than usually predicted by stellar models in the literature. As extensively discussed by \citet{Marigo_etal_20}, this feature is of key importance to explain the IFMR kink around $\Mi\simeq 2\, \Msun$.

\subsection{A Calibrated $\lambda$-Law at Solar Metallicity}
\label{ssec_lambda}
Taking advantage of the fact that our mass-loss prescription for carbon stars is sensitive to the surface carbon and oxygen abundances, we have the opportunity to introduce a methodology to calibrate the efficiency parameter, $\lambda$, of the 3DU, using the IFMR as an observable reference . 
The details of the procedure have already been illustrated in Section~\ref{sec_methods}.
We only recall that, compared to the approach used in  \citep{Marigo_etal_20, Kalirai_etal_14}, in this study we make a substantial improvement and include the dependence of $\lambda$ on the main stellar parameters
($\Mcmin,\, \Mc,\, \Menv$), based on  complete TP-AGB models \citep{Straniero_etal_03}.

\begin{figure}[H]
\centering
\includegraphics[width=6.5cm]{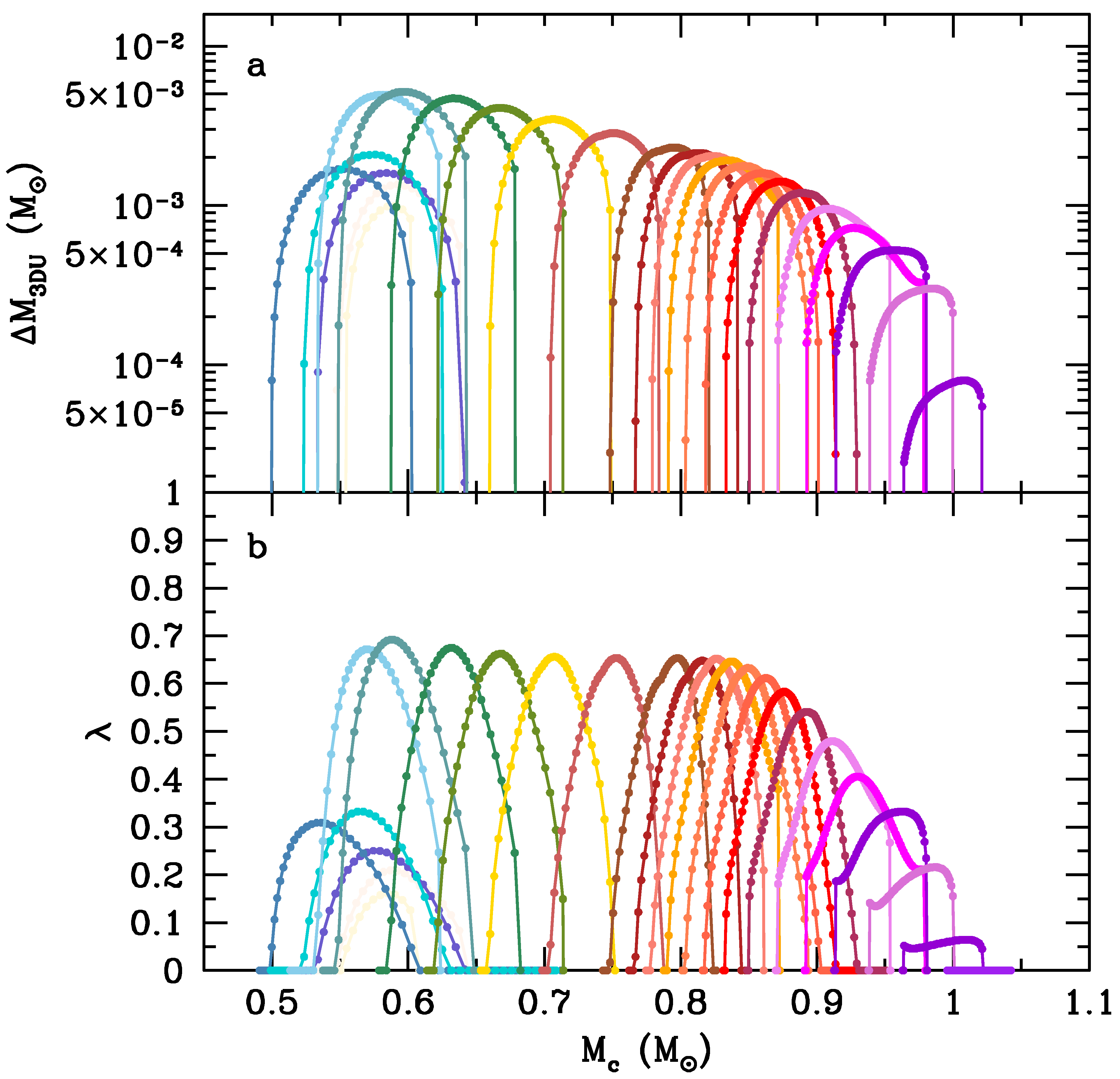}
\includegraphics[width=6.15cm]{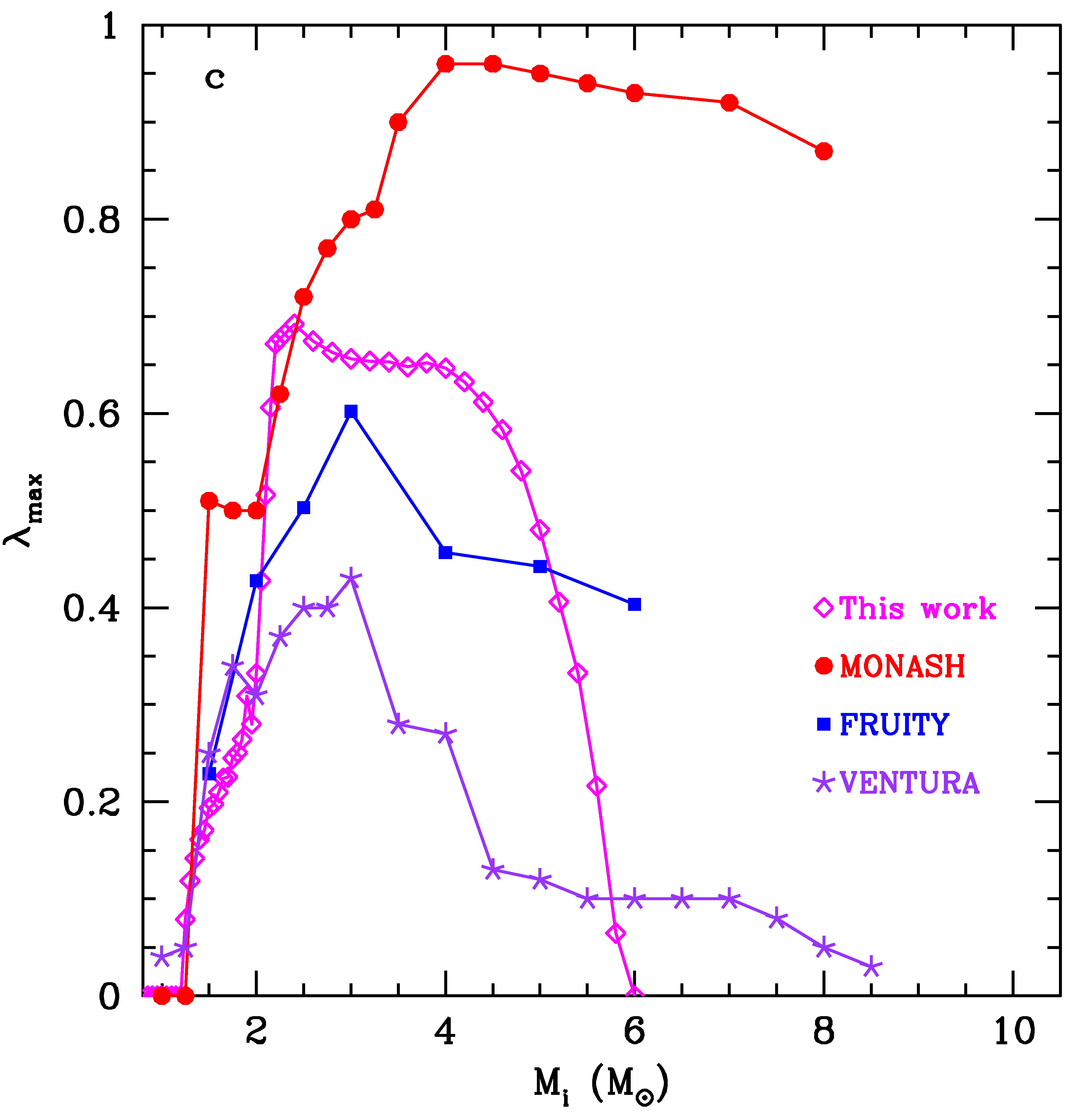}
\caption{Properties of the third dredge-up in TP-AGB stars at solar metallicity.
The IFMR-based calibration of the dredged-up mass \Mdup\ (\textbf{a}), and efficiency $\lambda$ (\textbf{b}), as a function of the current core mass.
\texttt{COLIBRI} TP-AGB models with solar metallicity cover the initial mass range 
$1.4 \lesssim \Mi/\Msun \lesssim 6.0$.
(\textbf{c}) The maximum efficiency of the 3DU as a function of the initial stellar mass, predicted from various authors, namely: This work [magenta diamonds]; (\cite{Ventura_etal_18}, Ventura, purple stars); (\cite{Karakas_Lugaro_18, Karakas_etal_14}, Monash, red circles); (\cite{Cristallo_etal_15}, FRUITY, blue squares).
\label{fig2}}
\end{figure}

The results of the 3DU calibration are shown in Figure~\ref{fig2} (left panel).
As expected, for each model, the efficiency of the 3DU first increases (following the growth of the  pulse strength), attains a maximum value, $\lambda_{\rm max}$, and then it decreases eventually reaching zero,  due to the reduction of the envelope mass by stellar winds. 

Some interesting aspects emerge from the analysis:
\begin{itemize}
\item Low-mass stars with $\Mi \lesssim 2\,\Msun$---those experimenting with the He-flash at the tip of the red giant branch---have a low 3DU efficiency, with $\lambda \lesssim 0.3$.
\item Intermediate-mass stars with $2 \lesssim \Mi \lesssim 6$---those that avoid electronic degeneracy in their helium cores---undergo a more efficient dredge-up. The $\lambda$ parameter  becomes more than twice as large
($\lambda_{\rm max} \simeq$ 0.6--0.7) in TP-AGB stars with $2 \lesssim \Mi/\Msun \lesssim 4.5$. \item At higher masses, $4.5 \lesssim \Mi \Msun \lesssim 6.0$, the efficiency decreases reaching zero at the largest \Mc.
\end{itemize}

{With respect to the latter point, we note that the decrease of $\lambda$ in the most massive TP-AGB stars follows from the constraint of obtaining sufficiently high final masses, in agreement with those measured for WDs members of young open clusters  \mbox{($\Mf >$ 0.9--1.0~$\Msun$)}.
From a physical point of view this could be explained by the fact that relatively lower efficiencies of the TDU are predicted at increasing core mass and hence shorter interpulse periods, as a consequence of the weaker thermal pulses \citep{Ventura_Dantona_08, Cristallo_etal_15}. Furthermore, following the analysis of \cite{Straniero_etal_14}
the combination of hot dredge-up (\cite{Goriely_Siess_04}) and HBB limits the occurrence of the third dredge-up in stars with $\Mi >$ 4--5 $\Msun$. By extrapolating the $\lambda$-calibration to higher masses ($\Mi > 6\, \Msun$), we could speculate that 3DU might be virtually absent in Super-AGB stars. However, this prediction requires a further observational verification, such as, for example, the absence or scarcity of heavy elements produced by the s-process in the atmosphere of these very luminous red super-giants.}

We can now compare our results with predictions from full TP-AGB models.
Panel \textbf{c} of Figure~\ref{fig2} shows the trend of the 3DU maximum efficiency, $\lambda_{\rm max}$, as a function of the initial mass of the star, as predicted by different sets of models.
We note that our calibration has a qualitative trend similar to the \texttt{FRUITY} and Ventura models: it increases with the initial mass, touches a maximum at $\Mi \approx$ 2.5--3.0 $\Msun$, and then drops at the highest masses.
On the other hand, when we quantitatively compare the results, substantial differences appear, with the largest deviations in $\lambda_{\rm max}$ showing up for $\Mi > 3\, \Msun$. 
For instance, taking  the $\Mi = 6 \,\Msun$ model, Ventura \citep{Ventura_etal_18} predicts $\lambda_{\rm max} \simeq 0.1$, \texttt{FRUITY} \citep{Cristallo_etal_15} yields  $\lambda_{\rm max} \simeq 0.4$ ,  while the Monash model of \citet{Karakas_etal_14} has $\lambda_{\rm max} \simeq 0.93$.
In general, our calibrated $\lambda_{\rm max}$  is closer to \texttt{FRUITY} and Ventura models, while it deviates sizeably  from the Monash models, being much lower for $\Mi > 2.5\,\Msun$.

\subsection{The IFMR: Data vs. Stellar Models}
\label{ssec_ifmr}
\begin{figure}[H]
\centering
\includegraphics[width=6.7cm]{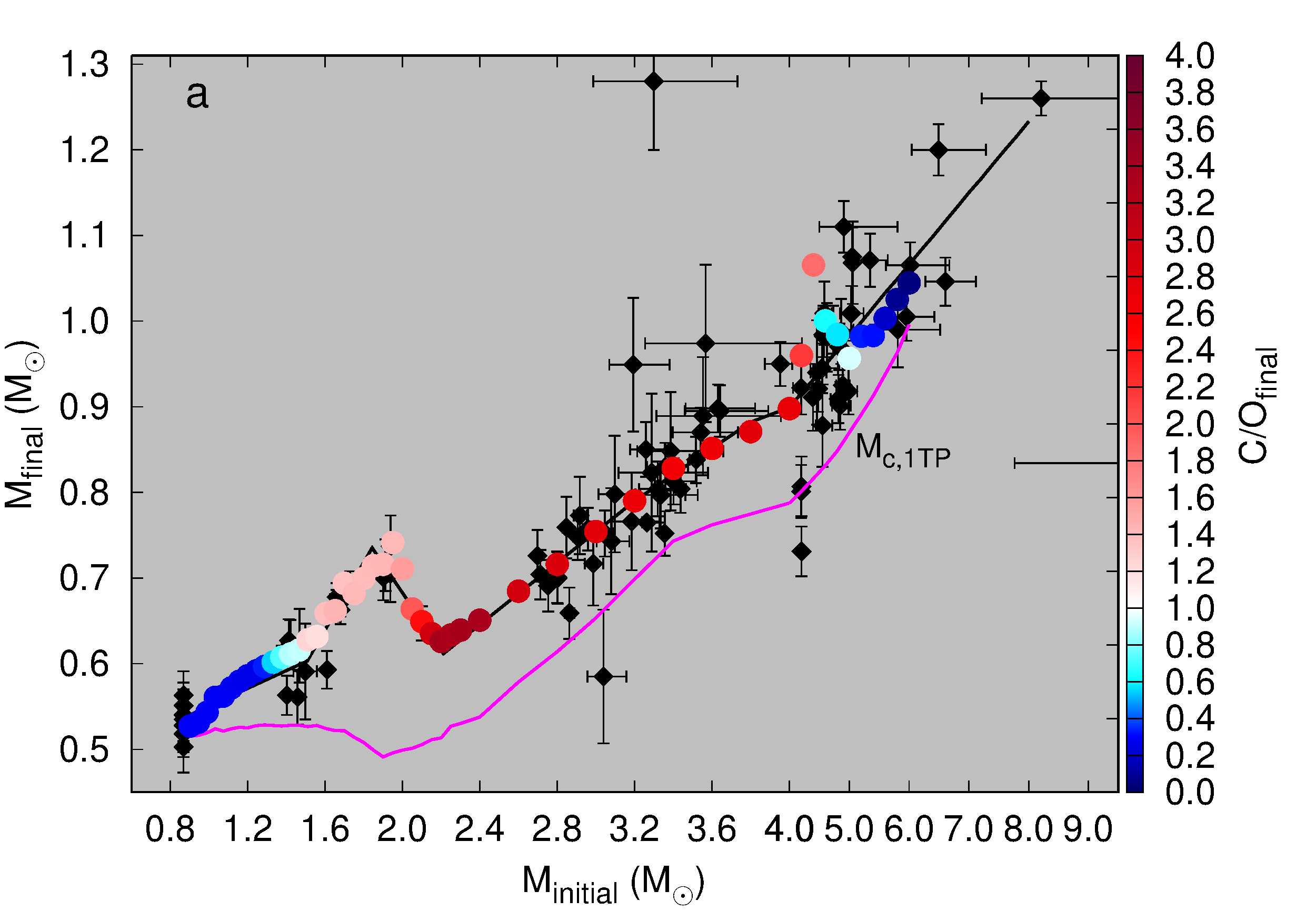}
\includegraphics[width=6.7cm]{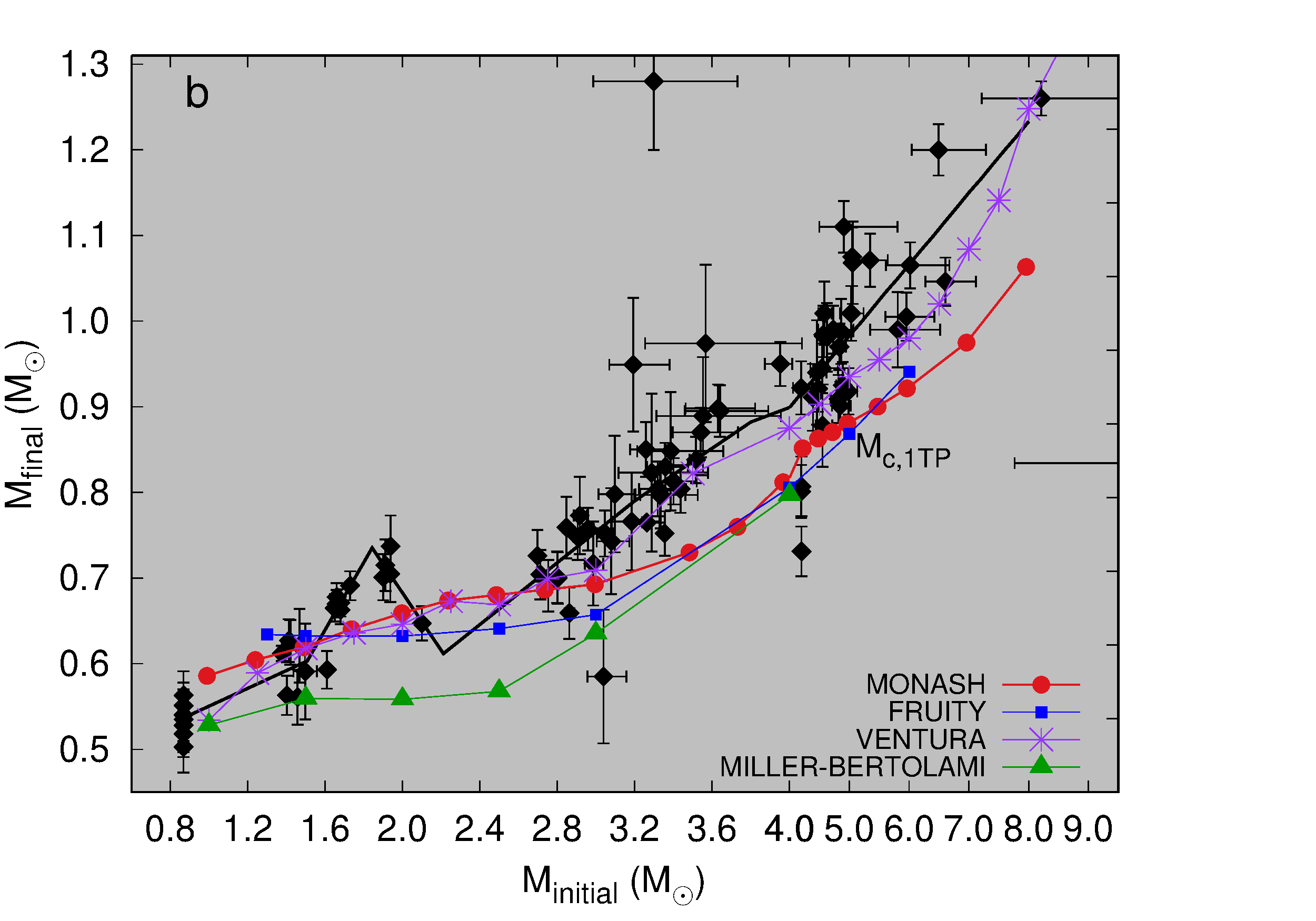}
\caption{
Comparison between theoretical and observed IFMR \citep{Marigo_etal_20, Cummings_etal_18}. The black solid line is a fit to the data. (\textbf{a}) Results of this work, obtained with the calibrated $\lambda$ values, color-coded as a function of the final \co\ (right color bar). The core mass at the first TP is also shown (magenta line). (\textbf{b}):~Predictions from various authors, namely:
(\citep{Ventura_etal_18}, Ventura, purple stars); (\citep{Karakas_Lugaro_18, Karakas_etal_14}, Monash, red circles); (\citep{Cristallo_etal_15},~FRUITY, blue squares); (\citep{Miller-Bertolami_16}, Miller--Bertolami, green triangles).
\label{fig3}}
\end{figure}

Figure~\ref{fig3} (panel \textbf{a}) 
shows the results of our $\lambda$-calibration at solar metallicity:
the semi-empirical IFMR is well reproduced by the \texttt{COLIBRI} models all over the range \linebreak\mbox{$0.9 \lesssim \Mi/\Msun \lesssim 6$}.

The minimum initial mass for carbon star formation is $\Mi \simeq 1.5\, \Msun$, in agreement with the observational evidence from Galactic open clusters \citep{Marigo_etal_22} and inference from studies on galactic enrichment and solar system formation \citep{Busso_atal_99}.
The WDs that populate the IFMR kink should be the progeny of low-mass carbon stars with $1.5 \lesssim \Mi/\Msun \lesssim 2.0$, which are characterized
by a shallow 3DU and a mild carbon enrichment.
The final \co\ lies in the range of 1.2--1.6 in these models.

In the domain of intermediate-mass stars ($2 \lesssim \Mi/\Msun \lesssim 4.2$),  at increasing \Mi\,  the 3DU is more efficient, and the surface composition becomes more enriched in carbon.  These models are expected to become carbon stars, with a final carbon-to-oxygen ratio of $2 \lesssim \co \lesssim 3.4$. In our grid, the $\Mi = 4.4\,\Msun$ model represents the most massive model that makes the transition to a carbon star and remains in this condition until the end of its evolution. This result is in line with 
the most massive carbon star found in open clusters:  BM IV 34, member of Haffner 14, has an estimated initial mass of $\simeq$3.4--4.0 $\Msun$.
At higher initial masses, ($4.4 \lesssim \Mi/\Msun \lesssim 6.0$), the models end the TP-AGB phase as O-rich stars. They all undergo hot-bottom burning, which becomes stronger at increasing \Mi. 
However, we note that there is a narrow mass interval, $4.6 \lesssim \Mi/\Msun \lesssim 4.8$, in which  the models experience a temporary C-rich phase, before being converted back to O-rich stars by HBB. This circumstance may leave a fingerprint in the IFMR, as discussed in Section~\ref{ssec_2kink}.

We can now examine the theoretical IFMRs published by different authors (panel~\textbf{b}  of Figure~\ref{fig3}). We see that models generally tend to underestimate the final masses of white dwarfs. The Miller--Bertolami relation \citep{Miller-Bertolami_16} is the one that most differs from the observations, while the Ventura relation \citep{Ventura_etal_18} is the one that comes closest to the data. 
No~model  reproduces the kink observed around $\Mi \simeq 2\,\Msun$. The reason is that none of the evolutionary calculations adopts 
a mass-loss prescription for carbon stars that is explicitly dependent on the carbon excess.

\subsection{A Second IFMR Kink at Higher Masses?}
\label{ssec_2kink}

Looking at Figure~\ref{fig3} (panel \textbf{a}) we note an interesting peculiarity: around  the transition mass between  carbon stars and  O-rich stars experiencing HBB, a second IFMR kink could appear. 
As can be seen, the $\Mi=4.4\,\Msun$ model ends its life as a carbon star and produces a white dwarf with $\Mf \simeq 1.06\, \Msun$,
while the $\Mi= 4.6, 4.8, 5.0\, \Msun$ models terminate the TP-AGB phase as  O-rich stars and leave progressively less massive white dwarfs, 
with $\Mf \simeq 1.00,0.98, 0.95\,\Msun$, respectively.
The reason lies in the different efficiency of stellar winds as a function of the surface chemical composition, in particular of the \co\ ratio. 
As to the $\Mi=4.4\, \Msun$ model, it is necessary that the 3DU enriches the envelope with enough carbon to start the dust-driven wind. In general, the greater the mass of the star, the stronger the inward gravitational force, and therefore the higher the minimum carbon threshold that must be reached. To meet this condition the star needs to experience a relatively large number of mixing episodes, and therefore, in the mean time, the masser of the core can grow.

\begin{figure}[H]
\centering
\includegraphics[width=6.8cm]{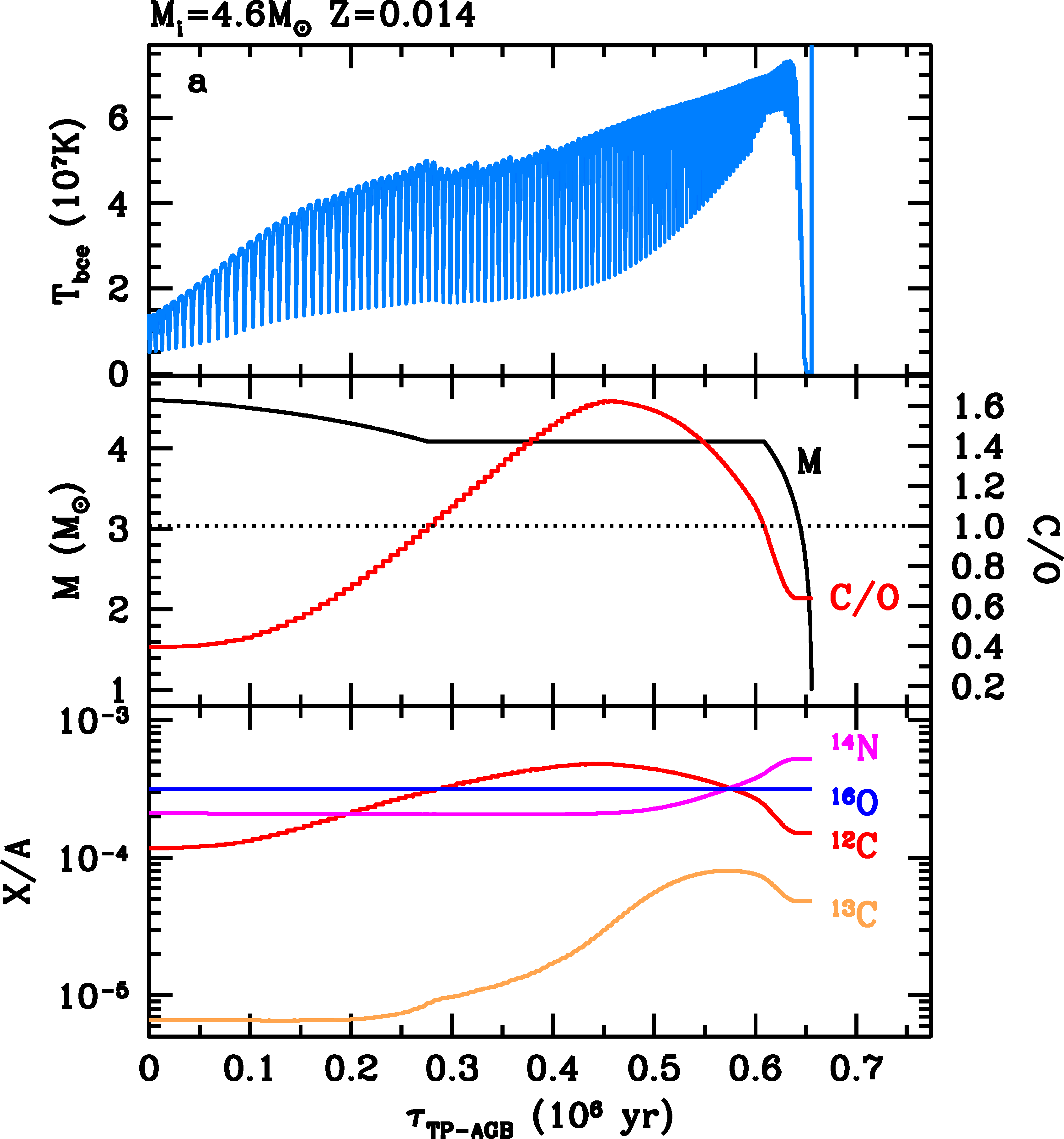}
\includegraphics[width=6.8cm]{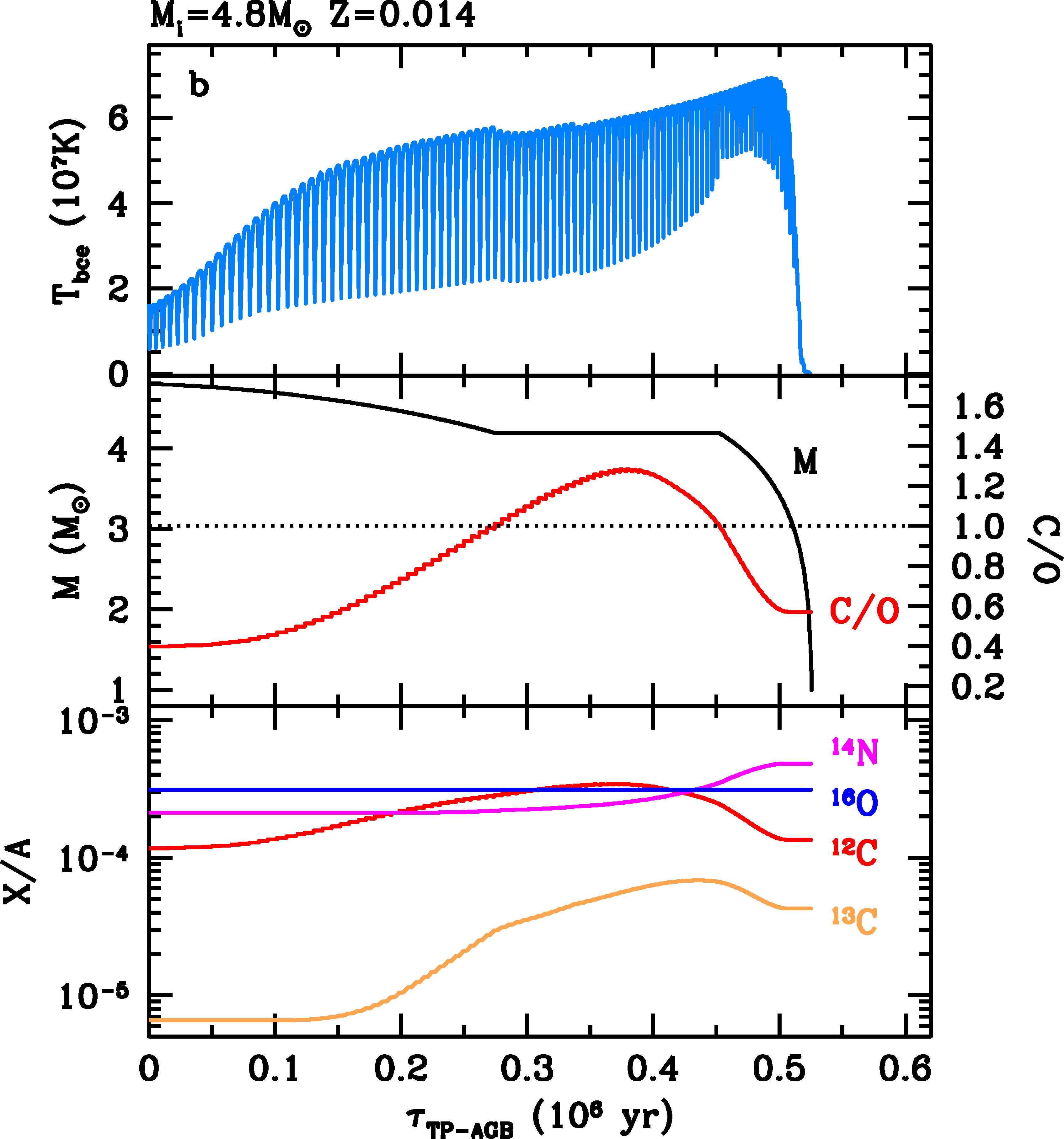}
\caption{Structural properties of the $Z=0.014$ models with  $\Mi=4.6\,\Msun$ (\textbf{a}), and $\Mi=4.8\,\Msun$ (\textbf{b}). Moving downward, the panels show the evolution during the whole TP-AGB phase of  the temperature, $T_{\rm bce}$, at the bottom the  convective envelope (top panel), the current mass being reduced by stellar winds, and the photospheric C/O (middle panel), the surface abundance (in molar fraction $X/A$, where $X$ is the mass fraction abundance, and $A$ denotes the mass number of the species) of a few elements (bottom panel). The horizontal dotted line corresponds to $\co=1$. Note that the during the temporary C-rich stages the evolution proceeds at almost constant mass. See the text for more details.
\label{fig4}}
\end{figure}

Moving at a higher stellar mass, the two models with $\Mi= 4.6, 4.8\, \Msun$, 
before leaving the AGB as  O-rich stars,  experience a transient and short C-rich phase, with $\co >1$. This circumstance can be appreciated in Figure~\ref{fig4}.
The  $\Mi= 4.6\, \Msun$ model is massive and bright enough to experience HBB, as well as a modest 3DU. Both processes can affect the surface \co\ in opposite ways: The 3DU which makes it grow instantaneously at each dredge-up episode, while HBB tends to lower it during the brighter phases of the interpulse periods.
Initially the competition is won by the 3DU, so that the star enters the C-star domain after about 40 TPs. During these C-rich phases, lasting $\simeq$0.33 Myr, the mass-loss rate undergoes a substantial slowdown for the reasons already discussed, and the evolution proceeds at an almost constant mass (middle panel). 

In the mean time, the growth of the core mass together with the increase in luminosity make the temperature of the deeper layers of the convective envelope increase. HBB becomes stronger and the surface \co\ ratio, after reaching a maximum of about 1.62, begins to decline due to the conversion of $^{12}$C and $^{13}$C into $^{14}$N by the CN cycle (middle panel).
At the $102$ {nd} TP the \co\ falls below one and the mass-loss rate switches to the \citet{Bloecker_95} prescription, which depends significantly on luminosity.  At this stage, $L \simeq 34\,000\,\Lsun$ and the mass-loss rate increases suddenly, leading to a rapid termination of the TP-AGB phase with a final mass $\Mf \simeq 1.00\,\Msun$.

The $\Mi=4.8\,\Msun$ model experiences a similar evolution, but being more massive and brighter, the HBB process is more efficient and therefore the duration of the C-rich phase, when mass loss undergoes a stop, is shorter ($\simeq$0.18 Myr). As a consequence, the final mass is smaller, $\Mf\simeq 0.98\,\Msun$.
Finally, starting from the $\Mi = 5\, \Msun$ model and upwards in mass, the C-rich transient phase disappears as HBB dominates, and therefore the theoretical IFMR  resumes an increasing monotonic trend.

In short, our study indicates that the application of a mass-loss prescription dependent on carbon excess may produce two kinks in the IFMR of white dwarfs:
\begin{enumerate}
    \item A first kink starts near the minimum mass for star formation at carbon, peaking at the transition mass between low- and intermediate-mass stars, covering approximately the $1.6\lesssim \Mi/\Msun \lesssim 2.2$ range;
\item A second kink shows up around the transition mass between the highest-mass carbon stars and those that remain O-rich due to the HBB process. Our models indicate a range $4.2\lesssim \Mi/\Msun \lesssim 4.8$, but the exact prediction may change depending on the
assumed mass-loss prescriptions, and on the efficiencies of the 3DU and HBB (hence on the adopted convection theory and/or mixing-length parameter). {A thorough study that aims to explore the fine details of the second IFMR kink is postponed for a future work.}
\end{enumerate}

While there is observational evidence for the existence of the first kink \citep{Marigo_etal_20, Marigo_etal_22}, the situation is less clear for the second kink. We only note  that the data in Figure~\ref{fig3} show a large dispersion in WD masses of between 4.5 \Msun and 6 \Msun, with some white dwarfs reaching $\Mf \simeq 1.1 \Msun$ at $\Mi \simeq 5\,\Msun$.
Were the second massive kink confirmed, it could serve as a powerful calibrator of the HBB efficiency, which is currently very uncertain in AGB star modelling.
{At the same time,  we should also recall that the scatter in WD masses at large \Mi\ may also be attributed to a dispersion of initial rotational velocities and convective core overshoot in the evolutionary phases prior the AGB (\citep{Cummings_etal_19}, see the discussion in). Both processes in main-sequence stars, rotational mixing and convective overshoot, lead to creating more massive cores and hence more massive WDs.}

\section{TP-AGB Lifetimes and Chemical Ejecta}
\label{sec_tau}
\begin{figure}[H]
\centering
\includegraphics[width=6.5cm]{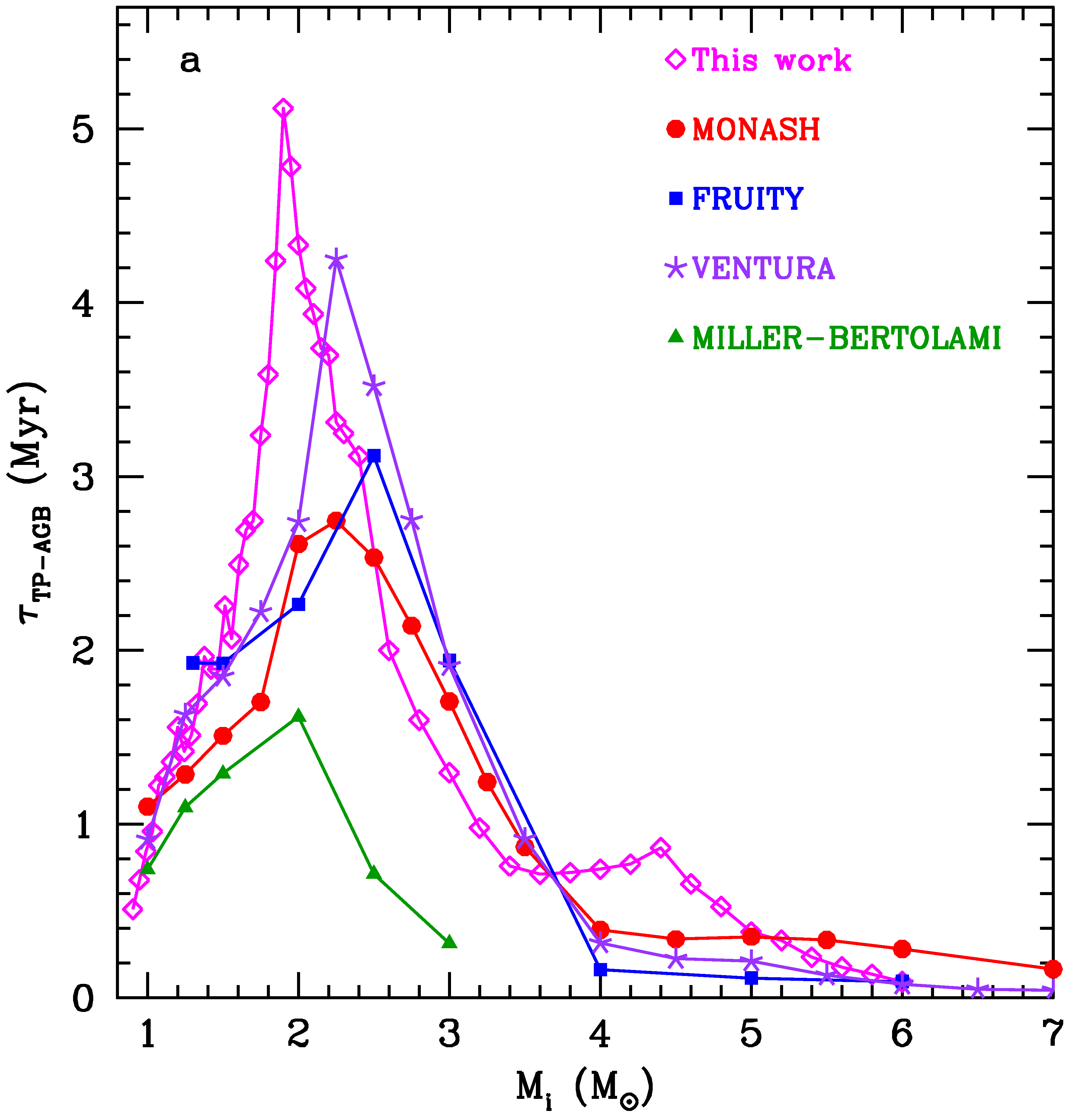}
\includegraphics[width=6.9cm]{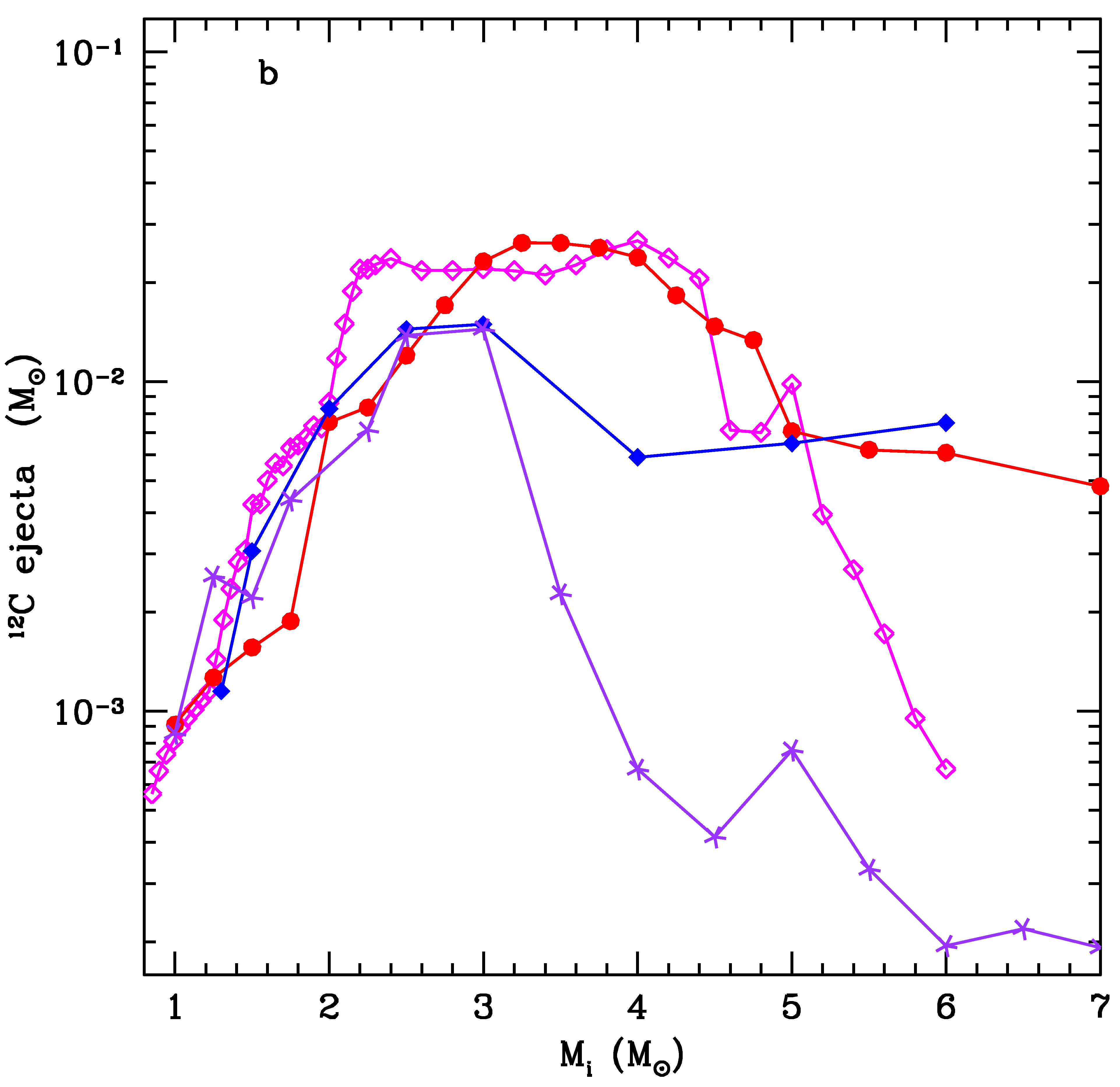}
\caption{Comparison of TP-AGB models at solar metallicity, from various sources. References and symbols are the same as in Figure~\ref{fig3}.
(\textbf{a}) TP-AGB phase lifetime as a function of the stellar initial mass.
(\textbf{b}) $^{12}$C ejecta contributed by TP-AGB stars as a function of the stellar initial mass. 
\label{fig5}}
\end{figure}

The $\lambda$-calibration affects the entire TP-AGB evolution. Here we focus on lifetimes and chemical yields.
Figure~\ref{fig5} (panel \textbf{a}) compares the duration of the TP-AGB phase, as a function of the initial mass for solar metallicity. We find that all models predict $\tau_{\rm TP-AGB}$ to be an initially increasing function with \Mi\, reaching a peak, and then decreasing to larger masses.
The position of the peak lies in the range $1.8\lesssim \Mi/\Msun \lesssim 2.5$, while its height, that is, the maximum TP-AGB lifetime, can vary by several factors, from about 1.6 Myr \citep{Miller-Bertolami_16} to $\simeq$5.2 Myr (this work).

We note that the relation $\tau_{\rm TP-AGB}(\Mi)$ predicted by our calibration shows a primary peak of $\simeq$5.2 Myr at $\Mi \simeq  1.8\, \Msun$, and a  smaller peak of $\simeq$0.9 Myr for $\Mi = 4.4\,\Msun$. The latter is produced by the temporary phase C-rich
already discussed in Section~\ref{ssec_2kink}, which is responsible for the second kink in the theoretical IFMR report (Figure~\ref{fig3}, panel \textbf{a}).
These predictions should be tested through population synthesis simulations to be compared with observational data (star counts, luminosity functions), similarly to the studies of \mbox{\citet{Pastorelli_etal_19, Pastorelli_etal_20}} for the Magellanic Clouds.

The number of dredge-up events, their efficiency and the total duration of the TP-AGB phase impact on chemical yields. In panel \textbf{b}  of Figure~\ref{fig5} we compare the $^{12}$C ejecta calculated by different authors. As can be seen, for $\Mi \lesssim 2\, \Msun$  differences exist but are relatively moderate, while large discrepancies among the authors occur for $\Mi > 2,\,\Msun$, due to different efficiencies of 3DU, HBB and different mass loss prescriptions adopted in the computations.
A more in-depth discussion of chemical yields, in light of our new models, will be carried out in a future study.

\section{Concluding Remarks}
\label{sec_end}
Benefiting from recent systematic studies of the IFMR \citep{Cummings_etal_18, Marigo_etal_20} and of state-of the art mass loss prescriptions for carbon stars \citep{Mattsson_etal_10, Eriksson_etal_14}, in this work we have introduced a methodology to calibrate the efficiency of the 3DU at solar metallicity, as a function of the parameters \Mcmin, \Mc, and \Menv.
Our calibration indicates that low-mass stars are characterized by low $\lambda\lesssim 0.3$, while the efficiency increases significantly in the class of intermediate-mass stars, and eventually decreases in massive AGB stars.
We emphasize that the calibration depends on the mass loss used in the models, which underlines the utmost importance of having reliable and up-to-date mass loss prescriptions to follow the TP-AGB phase. 

Our models are able to reproduce well the IFMR kink  around $\Mi \simeq 2,\Msun$, first identified by \citet{Marigo_etal_20}.
We also suggest the possible existence of a secondary kink, located in the approximate range $4\lesssim \Mi/\Msun \lesssim 5$, at the transition between the most massive carbon stars and those that remain O-rich due to the onset of HBB. This feature needs further theoretical and observational investigation to be confirmed or rejected.

{Finally, we remark again that these results apply to solar-like metallicities. In fact, one should consider that, in principle, the IFMR depends on metallicity (see, for example, \citep{Pastorelli_etal_19}) since the efficiencies of both the 3DU and AGB winds are affected by the envelope chemical composition. We will investigate this important aspect in a follow-up study.}



\funding{This research was funded by the ERC Consolidator Grant, Project STARKEY, grant agreement n. 615604, and from PRD 2021, University of Padova.}

\dataavailability{In this study we made use of data available through the FRUITY public website at \url{http://fruity.oa-teramo.inaf.it/}.} 

\conflictsofinterest{The author declares no conflict of interest. 
The funders had no role in the design of the study; in the collection, analyses, or interpretation of data; in the writing of the manuscript, or in the decision to publish the~results.} 

\abbreviations{Abbreviations}{
The following abbreviations are used in this manuscript:\\

\noindent 
\begin{tabular}{@{}ll}
IFMR & initial-final mass relation \\
TP-AGB & thermally pulsing asymptotic giant branch\\
TP & thermal pulse \\
WD & white dwarf\\
3DU & third dredge-up\\
HBB & hot-bottom burning \\
\Mi\ & initial mass \\
\Mf\ & final mass \\
\Mc\ & core mass \\
\Mcmin & minimum core mass for the  3DU \\
\Menv\ & envelope mass \\
$T_{\rm bce}$ & temperature at the base of the convective envelope \\
\co\ & surface carbon-to-oxygen ratio \\
\cminuso\ & carbon excess with respect to oxygen \\
\end{tabular}
}


\begin{adjustwidth}{-\extralength}{0cm}

\reftitle{References}



\end{adjustwidth}

\end{document}